\documentclass[aps,prev,twocolumn,preprintnumbers,floatfix,nofootinbib]{revtex4-1}
\pdfoutput=1

\newcommand{\be}{\begin{equation}}
\newcommand{\ee}{\end{equation}}
\newcommand{\bea}{\begin{eqnarray}}
\newcommand{\eea}{\end{eqnarray}}

\usepackage[utf8]{inputenc}
\usepackage{graphics}
\usepackage{mathrsfs}
\usepackage{amsmath, amsthm, amssymb}
\usepackage{color}
\usepackage{epstopdf}
\usepackage[pdftex]{graphicx}
\usepackage{amssymb}
\usepackage{verbatim}
\usepackage{hyperref}
\usepackage{enumerate}
\usepackage{float}
\usepackage[caption = false]{subfig}
\usepackage[normalem]{ulem}

\begin{document}

\title{Strong Constraints on Fuzzy Dark Matter from Ultrafaint Dwarf Galaxy Eridanus II}

\author{David J. E. Marsh}
\email{david.marsh@uni-goettingen.de}
\author{Jens C. Niemeyer}
\email{jens.niemeyer@phys.uni-goettingen.de}

\vspace{1cm}
\affiliation{Institut f\"{u}r Astrophysik, Georg-August Universit\"{a}t, Friedrich-Hund-Platz 1, D-37077 G\"{o}ttingen, Germany}

\begin{abstract}

The fuzzy dark matter (FDM) model treats DM as a bosonic field with astrophysically large de Broglie wavelength. A striking feature of this model is $\mathcal{O}(1)$ fluctuations in the dark matter density on time scales which are shorter than the gravitational timescale. Including for the first time the effect of core oscillations, we demonstrate how such fluctuations lead to heating of star clusters, and thus an increase in their size over time. From the survival of the old star cluster in Eridanus II we infer $m_a\gtrsim 0.6\rightarrow 1\times 10^{-19}\text{ eV}$ within modelling uncertainty if FDM is to compose all of the DM, and derive constraints on the FDM fraction at lower masses. The subhalo mass function in the Milky Way implies $m_a\gtrsim 0.8\times 10^{-21}\text{ eV}$ to successfully form Eridanus II. The region between $10^{-21}\text{ eV}$ and $10^{-20}\text{ eV}$ is affected by narrow band resonances. However, the limited applicability of the diffusion approximation means that some of this region may still be consistent with observations of Eridanus II. 

\end{abstract}

\maketitle
%%%%%%%%%%%%%%%%%%%%%%%%%%%%%%%%%%%%%%%%%%%%%%%%%%%%%%%%%%%%%%%%
%%%%%%%%%%%%%%%%%%%%%%%%%%%%%%%%%%%%%%%%%%%%%%%%%%%%%%%%%%%%%%%%

A wide variety of astrophysical observations require the existence of non-baryonic dark matter (DM)~\cite{2005PhR...405..279B,Aghanim:2018eyx,2015NatPh..11..245I,pdg}. At two extreme ends of the model space lie primordial black holes (PBHs) with masses as large as $M_{\rm PBH}\approx 10 M_\odot$, and ``fuzzy'' DM (FDM) composed of particles (possibly axions) as light as $m_a\approx 10^{-22}\text{ eV}$~\cite{hu2000,2014MNRAS.437.2652M,2016PhR...643....1M,Hui2017}. The fraction of DM allowed in heavy PBHs is severely constrained by the dynamics of stars in ultrafaint dwarf galaxies (UFDs)~\cite{Brandt2016,Li2016}. Two body relaxation and gravitational scattering between PBHs leads to heating of stars in the DM potential. This causes star clusters to grow in size on time scales incompatible with their observed sizes and ages, excluding a range of the PBH DM parameter space. In the following we will show that, somewhat remarkably, the very same observations of old star clusters in UFDs place strong constraints on FDM parameter space.  

FDM is modelled as a coherent bosonic field, $\phi$. In the minimal non-interacting case the potential is $V(\phi)=m_a^2\phi^2/2$, and the field coherently oscillates in the minimum. This coherence leads to fluctuations on two distinct time scales. Firstly, relativistic Compton scale fluctuations of order $m_a^{-1}\approx m_{22}^{-1}\text{ month}$, where $m_{22}=m_a/10^{-22}\text{ eV}$, lead to pressure perturbations, and in turn metric fluctuations and can be searched for by a variety of techniques ~\cite{2014JCAP...02..019K,Blas:2016ddr,Aoki:2016kwl,DeMartino:2017qsa,Boskovic:2018rub}. They also underlie methods of direct detection of FDM~\cite{Abel:2017rtm}. In the present work we neglect Compton fluctuations since the time scale is not relevant to the dynamics of star clusters.

Fluctuations also occur on the de Broglie scale, $\lambda_{\rm dB}=2\pi/m_a v$, with oscillation period $\tau_{\rm osc}=2\pi/m_a v^2$ (we use units $\hbar=c=1$). In linear theory, these fluctuations manifest as the FDM Jeans scale~\cite{khlopov_scalar}, which suppresses structure formation relative to cold DM (CDM). This drives cosmological constraints on FDM~\cite{Hlozek:2017zzf,2015MNRAS.450..209B,2016ApJ...818...89S,2016JCAP...04..012S,2017PhRvD..95h3512C,Irsic:2017yje,Armengaud:2017nkf,Nori:2018pka}, leading to the bound $m_a\gtrsim 10^{-22}\rightarrow 10^{-21}\text{ eV}$ depending on the data and modelling. In terms of the halo mass function, numerical~\cite{2016ApJ...818...89S,2017PhRvD..95h3512C} and semi-analytical~\cite{2014MNRAS.437.2652M,2017MNRAS.465..941D} calculations predict that the abundance of halos in FDM is severely reduced relative to CDM for masses less than $M_{\rm cut}\approx 3\times 10^8 m_{22}^{-3/2} M_\odot$.

Inside DM halos the de Broglie fluctuations are observed in simulations as granular structure in the outer halo resulting from wave interference~\cite{2014NatPh..10..496S,Veltmaat2018}. It is the central insight of Ref.~\cite{Hui2017} that these fluctuations can be treated statistically as short-lived quasiparticles, and lead to heating effects and relaxation in a similar way to PBHs. The relaxation time is estimated as:
\be
\frac{t_{\rm relax}}{10^{10}\text{ yr}} \sim m_{22}^3\left( \frac{v}{100\text{ km s}^{-1}}\right)^2\left( \frac{r}{5\text{ kpc}}\right)^4 \, .
\ee
The effect of FDM fluctuations on stellar dynamics in the Milky Way (MW) region has been investigated extensively, imposing constraints on the FDM mass of $m_a>0.6\rightarrow 1.5 \times10^{-22}\text{ eV}$ from the thickening of the disk~\cite{2018arXiv180904744C} and stellar streams~\cite{Amorisco:2018dcn} respectively.

FDM simulations also point to the existence of a central solitonic DM core on the de Broglie scale~\cite{2014NatPh..10..496S}. In zoom-in simulations of FDM galaxies~\cite{Veltmaat2018} it is observed that the central soliton is not stationary, as was previously thought, but undergoes quasi-coherent oscillations in its central density, with a relative amplitude of $\mathcal{O}(30\%)$ and period $\mathcal{O}(\tau_{\rm osc})$. The present work presents the first study of the effect of core oscillations on stellar dynamics.

FDM solitonic cores are observed to form in simulations of dwarf galaxies with $M\approx 10^{10}M_\odot$ when $m_a\approx 10^{-22}\text{ eV}$. They form by direct collapse almost instantaneously when the halo virialises. For larger FDM masses, however, it is not clear whether soliton formation in dwarf galaxies will proceed in the same way, since the length scales involved are much longer than the de Broglie wavelength. The time scale for soliton formation by wave condensation increases at larger particle masses~\cite{Levkov:2018kau}, and thus solitons may not have had time to form in all halos for all FDM masses. 

Assuming it forms, the central soliton has the density profile of the ground state of the Schr\"{o}dinger-Poisson equation. The solution $\rho_{\rm sol}(r)$ is a one parameter family described by the core radius, $r_c$, and has a flat central density, $\partial_r\rho|_{0}=0$. The soliton mass within the core radius is observed to follow a scaling relation with the host halo mass, which at redshift $z=0$ is given by~\cite{2014PhRvL.113z1302S,Du:2016aik}:
\begin{equation}
    M_{\rm sol}=\frac{M_0}{4}\left(\frac{M_h}{M_0}\right)^{1/3}\, ,
    \label{eqn:core_halo_mass}
\end{equation}
where the scale $M_0\approx 4.4\times 10^7 m_{22}^{-3/2} M_\odot$ is approximately the Jeans mass. The relation Eq.~\eqref{eqn:core_halo_mass} can be used to fix $r_c$ in terms of $M_h$:
\begin{equation}
    r_c = 740 \left(\frac{m_a}{10^{-21}\text{ eV}}\right)^{-1}\left( \frac{M_h}{10^7M_\odot}\right)^{-1/3}\text{ pc}\, .
    \label{eqn:core_halo_radius}
\end{equation}
The core-halo mass relation constrains FDM based on galactic rotation curve observations~\cite{Bar:2018acw}. 

The central soliton has some favourable consequences, e.g. its stabilising effect on the cold clump in Ursa Minor~\cite{2012JCAP...02..011L}, a possible explanation for cored density profiles in dSphs~\cite{2014NatPh..10..496S,2015MNRAS.451.2479M,2017MNRAS.468.1338C,Gonzales-Morales:2016mkl} and UFDs~\cite{2016MNRAS.460.4397C}, help alleviating the ``too big to fail'' problem~\cite{2014MNRAS.437.2652M,Robles:2018fur}, and an explanation for excess mass in the centre of the MW~\cite{2018arXiv180708153D}%(though the cusp-core problem in $M\approx 10^{11}M_\odot$ galaxies is exacerbated~\cite{Robles:2018fur}). 
These observations, as well as other hints from the small-scale structure of DM~\cite{Hui2017,2014MNRAS.437.2652M,2018ApJ...862..156L}, point to a preferred FDM mass $m_{22}=\mathcal{O}({\rm few})$.

\emph{Eridanus II} (Eri II) is a UFD with a centrally located star cluster~\cite{Li2016,2016ApJ...824L..14C}. Eri II is located at a distance of 370 kpc from the centre of the MW. The mass within the half-light radius is estimated as $M_{\rm EII}=1.2^{+0.4}_{-0.3}\times 10^7\,M_\odot$, 1D velocity dispersion $\sigma_v=6.9^{+1.2}_{-0.9} \text{ km s}^{-1}$, and central DM density $\rho_{\rm DM}=0.15\, M_\odot\text{ pc}^{-3}$. The central star cluster has a half light radius $r_h=13\text{ pc}$, age $T_{\rm EII}=3\rightarrow 12\text{ Gyr}$ and mass $M_\star=2000\, M_\odot$. These values have been shown to be consistent with the expected dynamical evolution in the presence of a DM core, disfavouring a cuspy DM profile~\cite{Contenta2018}.

\begin{figure}
\includegraphics[width=\columnwidth]{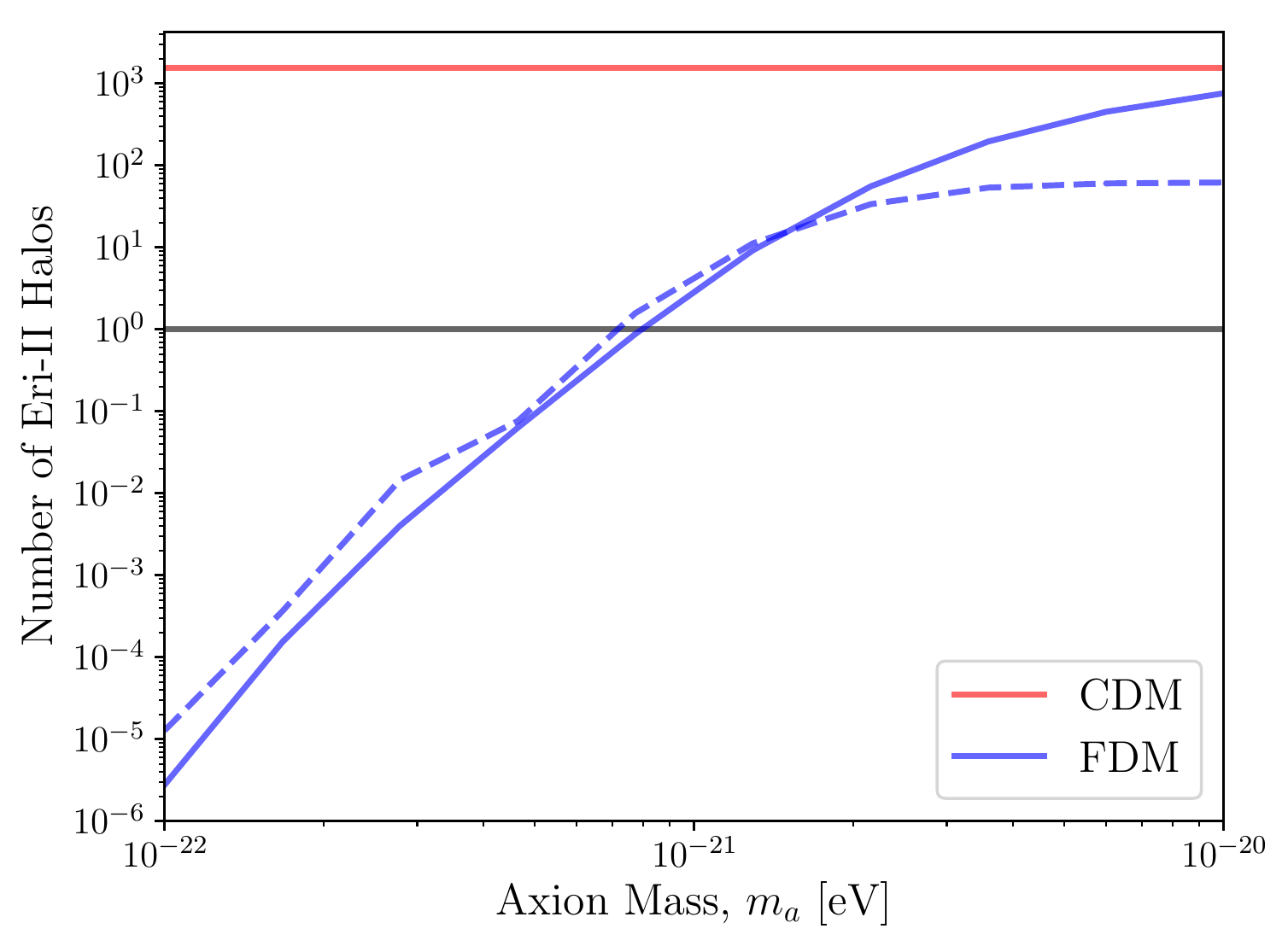}
\caption{Number of subhalos in the range of the mass of Eri II as a function of FDM mass $m_a$. Solid: from merger trees with modified barrier (with tidal stripping: present day half-light mass); dashed: sharp-$k$ filter (no tidal stripping: estimated pre-infall mass). We demand FDM produce at least one subhalo (black horizontal line). The horizontal red line shows the CDM prediction with tidal stripping.} 
\label{fig:subhalo}
\end{figure}

We can use these basic properties of Eri II to assess the relevant FDM scales. The total number of MW subhalos in the 2$\sigma$ range around $M_{\rm EII}$ ($M_{\rm low}=4\times 10^6 M_\odot$, $M_{\rm up}=2\times 10^7 M_\odot$) is
\begin{align}
    n_{\rm EII}(m_a) = \int_{M_{\rm low}}^{M_{\rm up}}{\rm d}\ln M \frac{{\rm d}n_{\rm sub}(m_a)}{{\rm d}\ln M} \, ,
\end{align}
where ${\rm d}n_{\rm sub}/{\rm d}\ln M$ is the subhalo mass function (see Fig.~\ref{fig:subhalo}). We estimate the FDM subhalo mass function with the fits of Ref.~\cite{XiaolongThesis}, which uses the methods of Refs.~\cite{Du:2016zcv,Du:2018zrg,2014MNRAS.437.2652M,Hui2017} applied to merger trees, and includes a model for tidal stripping~\cite{2012NewA...17..175B}. The exclusion on $m_a$ implied by the existence of Eri II is found by setting $n_{\rm EII}(m_a)=1$, and gives the approximate bound $m_a\gtrsim 8\times 10^{-22}\text{ eV}$ if FDM is all of the DM. As a comparison we also test the subhalo mass function of Refs.~\cite{Schneider:2014rda,2016JCAP...04..059S,2017JCAP...11..046M} computed using the sharp-$k$ filtering method~\cite{Schneider:2014rda}. The sharp-$k$ filtering model does not include stripping, and should be compared to the pre-infall mass of Eri II, $5\times 10^8 M_\odot$~\cite{Contenta2018}. The two models give comparable constraints on the FDM mass. When $m_a= 10^{-21}\text{ eV}$ Eri-II is a single core remnant (see also Ref.~\cite{Eby:2018zlv}). For larger values of $m_a$, Eri II will have a granular outer halo in addition to the core. 

The stability of the star cluster in Eri II can be taken to imply the existence of a DM core with radius $r_c\geq r_h$. %We estimate the FDM mass preferred by a core in Eri II by setting $\rho_{\rm sol}(r_h)=\rho_{\rm DM}$, which implies $m_a\approx 10^{-20}\text{ eV}$. This is significantly larger than the FDM mass required for cored profiles in UFDs Draco II or Triangulum III~\cite{2016MNRAS.460.4397C}, or dSphs Fornax and Sculptor~\cite{2015MNRAS.451.2479M,2017MNRAS.468.1338C} to be explained by the presence of a soliton. 
Assuming that the total mass of Eri II is given  by $M_{\rm EII}$, using Eq.~\eqref{eqn:core_halo_radius} with $M_h=M_{\rm EII}$ we can fix $r_c=r_h$ and solve for $m_a$ to find the highest possible FDM mass consistent with the star cluster residing within the soliton core, giving $m_a\approx 10^{-19}\text{ eV}$. For $m_a\lesssim 10^{-20}\text{ eV}$ the Eri II star cluster is guaranteed to be inside the soliton core. For $10^{-20}\text{ eV}\lesssim m_a\lesssim 10^{-19}\text{ eV}$ it is possible for the star cluster to lie either inside or outside the soliton within the (aproximate) observational uncertainty on the location of the star cluster.%, and theoretical uncertainty on the formation time of the soliton. 

\emph{Diffusion Approximation: Star Cluster Heating} 
Small fluctuations of the gravitational potential averaged over the orbital period increase the energy of stellar orbits (gravitational heating)~\cite{BinneyTremaine2008}. This effect can be computed in the diffusion approximation provided that the stellar orbital period, $\tau_{\rm orb}$, is long compared to the timescale of the fluctuations which, in the core, is set by the period of stochastic oscillations $\tau_{\rm osc}$.
%The diffusion approximation applies for outer halo fluctuations and stochastic core oscillations as long as $\tau_{\rm osc}$ is less than the stellar orbital period, $\tau_{\rm orb}$. 
The typical oscillation frequency is $\omega=m_a \sigma_{3D}^2$, with $\sigma_{3D}=\sqrt{3}\sigma_{1D}$. Taking the stellar period to be the Keplerian period we find:  
\begin{equation}
    \frac{\tau_{\rm orb}}{\tau_{\rm osc}} \sim \frac{m_a}{10^{-21}\text{ eV}}\, .
    \label{eqn:periods}
\end{equation}
We %assume that the density fluctuations have a shot noise distribution 
%produced by the granular interference structure of the scalar field, 
relate the spatial and temporal fluctuations with the dispersion velocity $v$ of the dark matter as $r = v t$. We neglect the intrinsic relaxation caused by the cluster stars to obtain an upper limit on the allowed amplitude of dark matter fluctuations. Two-body relaxation by cluster stars has been shown to naturally explain the observed radius of Eri II \cite{Contenta2018}.

%%%%%%%%%%%%%%%%%
\begin{figure*}
\includegraphics[width=1.7\columnwidth]{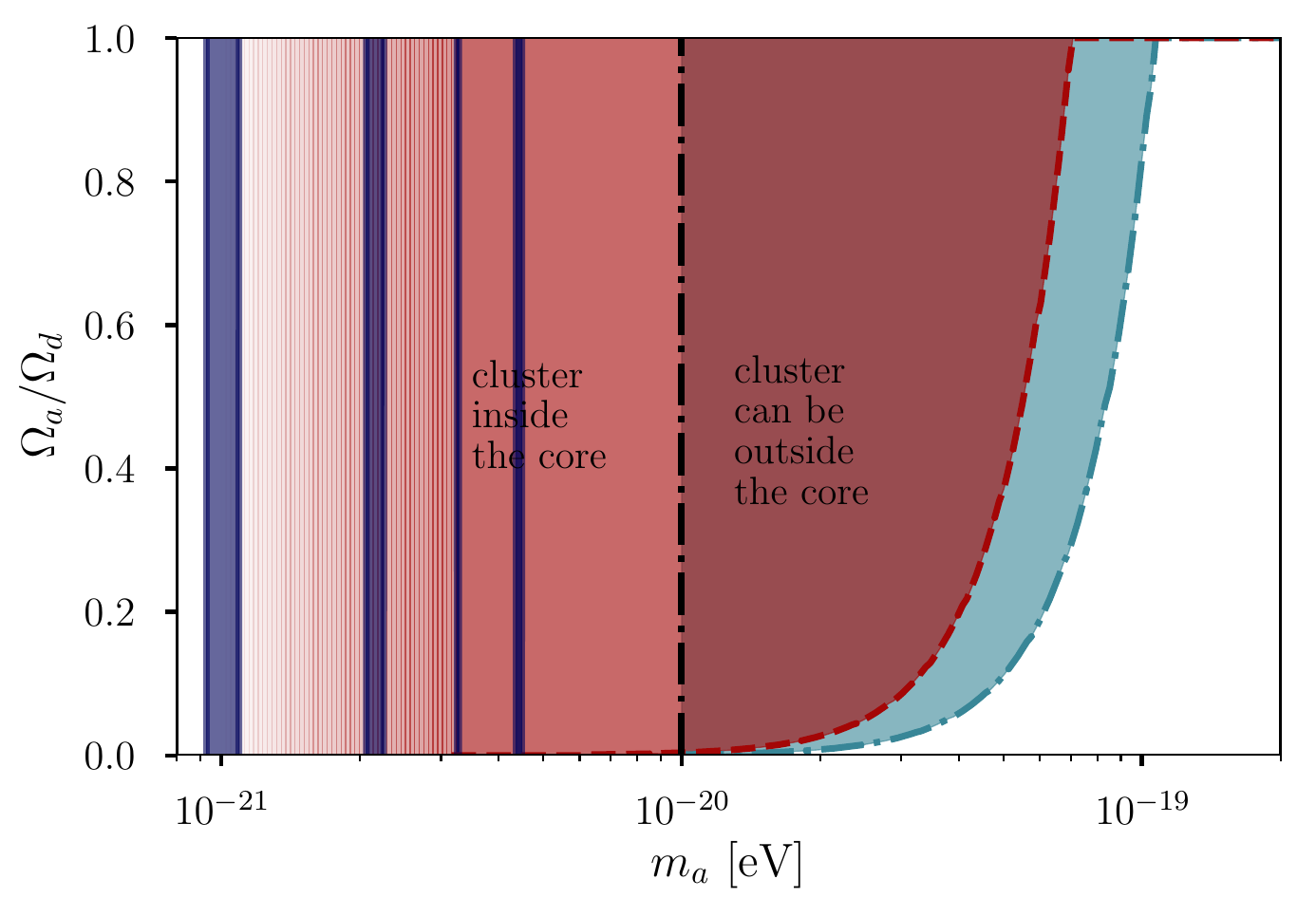}
\caption{FDM exclusions from the size and age of the Eri II star cluster. For $m_a\lesssim 10^{-20}\text{ eV}$ (vertical dashed line) the star cluster must be contained inside the soliton core and is affected by core oscillations with amplitude $\mathcal{C} = 0.3$ (red). For $m_a\gtrsim 10^{-20}\text{ eV}$ the star cluster may extend outside of the core and be subject to granular density fluctuations in the halo with $\mathcal{C} = 1$ (blue). Shaded regions are excluded by the diffusion approximation to heating of the star cluster. The validity of the diffusion approximation becomes questionable in the range $m_a=\mathcal{O}({\rm few})\times 10^{-21}\text{ eV}$, where the oscillation period and stellar period become similar (Eq.~\ref{eqn:periods}) (indicated by lighter shading in this region). Perturbation theory analysis of core oscillations excludes a series of narrow band resonances in the range $10^{-21}\text{ eV}\lesssim m_a\lesssim 5\times 10^{-21}\text{ eV}$. If $\Omega_a/\Omega_d=1$, Eri II does not form for $m_a\lesssim 0.8\times 10^{-21}\text{ eV}$. } 
\label{fig:plot1}
\end{figure*}
%%%%%%%%%%%%%%%

%These assumptions are well justified in the incoherent halo outside of the solitonic core. Inside the core, as long as the coherent fluctuation timescale is small compared to the typical orbital period of the stars, we can apply the same estimate for the gravitational heating rate by multiplying with a suppression factor for the fluctuation amplitude. 

We derive the gravitational heating rate produced by a fluctuating density field from the force correlation function, closely following Ref.~\cite{El-Zant2016} for the case of a turbulent baryon field~\footnote{The main difference to \cite{El-Zant2016} is that density fluctuations are dominated by the smallest scale in our case as opposed to the largest scale in theirs.}. The force correlation function is the Fourier transform of the force power spectrum, given by
\begin{equation}
    \langle F(0)F(r) \rangle = \frac{1}{2\pi^2} \int\, \mathcal{P}_F(k)\, \frac{\sin(kr)}{kr}\, k^2 dk
\end{equation}
assuming statistical isotropy. The force power spectrum produced by fluctuations of the gravitational potential, $\Phi$, in the volume $V$,
\begin{equation}
    \mathcal{P}_F(k) = V k^2 \langle \vert \Phi_k \vert^2 \rangle \,,
\end{equation}
is related to the power spectrum of density fluctuations
\begin{equation}
    \mathcal{P}_\delta(k) = V \langle \vert \delta_k \vert^2 \rangle \,,
\end{equation}
where $\delta_k$ are the Fourier components of the density contrast $\delta = \rho/\rho_0 - 1$, by the Poisson equation $k^2 \Phi_k = - 4\pi G \rho_0 \delta_k$.

We assume a $k$-independent shot noise density power spectrum, $\mathcal{P}_\delta \sim n^{-1}$, with $n \sim (l_c/2)^{-3}$ determined by the scalar field coherence length $l_c = 2\pi/k_c  \sim (mv)^{-1}$. In this case,
\begin{equation}
   \mathcal{P}_F(k) = (4 \pi G \rho_0)^2 \, \mathcal{P}_\delta \, k^{-2} 
\end{equation}
and
\begin{equation}
     \langle F(0)F(r) \rangle = \frac{C}{r} \int_{k_0}^{k_c}\, \frac{\sin(kr)}{k} \,dk = \frac{C}{r}\, \left.\mathrm{Si}\right\vert_{k_0}^{k_c}
\end{equation}
where $k_0$ corresponds to the largest fluctuation scale and $C = 8 G^2 \rho_0^2 \mathcal{P}_\delta$.

Following Ref.~\cite{El-Zant2016} (see also Ref.~\cite{Osterbrock1952}), we compute the velocity variance induced by the force fluctuations on the trajectory of a star in the cluster during the time $\tau$ as
\begin{align}
\label{eq:delv}
    \langle (\Delta v)^2 \rangle &= 2 \int_0^\tau \, (\tau-t)\,\langle F(0)F(t) \rangle \nonumber \,dt\\
            &= \frac{2}{v^2} \int_0^{v\tau}\, (v\tau-r)\,\langle F(0)F(r) \rangle \,dr \,,
\end{align}
where we used the dark matter velocity dispersion $v$ to relate the temporal fluctuations to the spatial ones as explained above. In the diffusion limit we demand that $\tau \gg \tau_{\rm osc}$, i.e. that the orbital period of the stars is greater than the fluctuation time scale. %This is always satisfied (XXX more details here or below? XXX).

In the limit $k_0 v\tau \ll 1$, Eq.~\eqref{eq:delv} evaluates to
\begin{align}
    \langle (\Delta v)^2 \rangle = \frac{2C\tau}{v} [&(k_c v\tau)^{-1}\left(1-\cos(k_c v\tau)\right)  \nonumber\\ 
    &+ k_c v\tau \,{}_2F_3\left(\frac{1}{2}, \frac{1}{2}; \frac{3}{2}, \frac{3}{2}, \frac{3}{2};-\left(\frac{k_c v\tau}{2}\right)^2\right) \nonumber \\
    &- \mathrm{Si}(k_c v\tau) ] \,,
\end{align}
where ${}_2F_3$ is the generalized hypergeometric function. Considering the diffusion limit $k_c v\tau \gg 1$, we can neglect the first term in square brackets, the last one asymptotes to $\pi/2$, and the middle one gives approximately $\pi/2 \,(\log(k_cv\tau) + 0.6)$. Together, we obtain 
\begin{equation}
\label{eq:delv2}
    \langle (\Delta v)^2 \rangle \simeq \frac{\pi C\tau}{v}\,\log(k_c v\tau)\,,
\end{equation}
where the logarithmic term can be identified with the Coulomb logarithm, i.e. the logarithm of the ratio of the largest and smallest relevant length scales of the system.

The relaxation time is defined as the time $\tau$ for which the induced velocity variance equals the mean square velocity of the stars $v_\ast^2$,
\begin{equation}
    t_\mathrm{relax} = \frac{v_\ast^2 v}{\pi C\,\log(k_c v\tau)}\,.
\end{equation}
Finally, the diffusion coefficient for the gravitational heating of the star cluster is given by \cite{BinneyTremaine2008}
\begin{align}
\label{eq:diffcoeff}
    D\left[\left(\Delta v\right)^2\right] &= \frac{v_\ast^2}{t_\mathrm{relax}} = \frac{\langle (\Delta v)^2 \rangle}{\tau} \nonumber \\
            &\simeq  \frac{8 \pi G^2 \rho_0^2 \mathcal{P}_\delta}{v}\,\log(k_c v\tau)\,.
\end{align}

It is interesting to compare Eq.~\eqref{eq:diffcoeff} with the corresponding diffusion coefficient for gravitational heating by MACHOs \cite{BinneyTremaine2008} applied to the Eri II star cluster by  Brandt~\cite{Brandt2016}. Replacing the MACHO mass in Brandt's Eq. (1) with the mass of granular quasiparticles \cite{Hui2017}, $m_\textrm{qp} = \rho_0\,(l_c/2)^3 = \rho_0\, \mathcal{P}_\delta$, we obtain the identical result for the heating rate up to a factor of $\sqrt{2}$ and the precise definitions of the Coulomb logarithm which are $\lesssim \mathcal{O}(10)$ in both cases. This demonstrates that the quasiparticle model for FDM and shot noise density fluctuations produced by interference patterns of the scalar field make equivalent predictions for the diffusion coefficient (see \cite{2018arXiv180907673B} for an in-depth discussion of diffusion coefficients in FDM scenarios).

Using $d v_\ast^2/dt = D$ and the virial theorem, one can find an equation for the growth of the star cluster radius \cite{Brandt2016}. The half-light radius $r_h$ evolves as
\begin{equation}
\label{eq:drhdt}
    \frac{d r_h}{dt} = {\cal C}\mathcal{F} \frac{8 \pi G \rho_0 \mathcal{P}_\delta }{v} \, \log(k_c v\tau) \, 
                 \left(\alpha \frac{M_\ast}{\rho r_h^2 } + 2 \beta r_h\right)^{-1} \,,
\end{equation}
where $\mathcal{F} = \Omega_a/\Omega_d$. For diffusion caused by the density granules in the outer halo we have $\mathcal{C}=1$, while for diffusion inside the core we take $\mathcal{C}=0.3$ to account for the amplitude of core density fluctuations found in simulations \cite{Veltmaat2018}~\footnote{Supernova feedback is known to enhance the amplitude of gravitational fluctuations and make them more stochastic. This is confirmed by simulations with FDM \cite{Veltmaat2019}. We consider the pure FDM case as a conservative lower limit.}. We use ten orbital periods of stars in the cluster to estimate $\tau$ in the Coulomb logarithm and set $\alpha = 0.4$, $\beta = 10$~\cite{Brandt2016}. To constrain the axion mass $m_a$ we impose that the time for $r_h$ to grow from 2 pc to 13 pc must be longer than the age of the cluster~\cite{Brandt2016}, 3 Gyr (an initial value of 1 pc reduces the limit on $m_a$ by $\sim 10 \%$).  The exclusions on $(m_a,\Omega_a/\Omega_d)$ are shown in Fig.~\ref{fig:plot1}.

\emph{Star Cluster Resonances} The star cluster evolution time scale caused by coherent density fluctuations inside the core can also be estimated using standard perturbation theory~\cite{goldstein1980classical}. The DM mass contained within the half-light radius is $M_{\rm DM}(r<r_h) \approx (4/3)\pi\rho_{\rm DM} r_h^2 = 330 M_\odot$, assuming the density is cored, giving $M_{\rm DM}<M_\star$, suggesting that the star cluster is self-bound. Consider a star of mass $m_\star$ on a Keplerian orbit with semi-major axis $a_0=r_h$ about the centre of mass of the star cluster, $V_0=-GM_\star m_\star/r=-k/r$, in terms of the action-angle variables $(w_i,J_i)$ in the limit $m_\star\ll M_\star$. The unperturbed Hamiltonian is:
\begin{equation}
H_0 =\frac{2\pi^2 m_\star k^2}{J_3^2}\, , 
\end{equation}
giving $w_3=t/\tau_{\rm orb}+{\rm const.}$ with $\tau_{\rm orb}=0.1 \text{ Gyr}$ is the Keplerian period. The semi-major axis $a_0=J_3^2/4\pi^2 m_\star k$. 

The size fluctuations of the solitonic core~\cite{Veltmaat2018} imply that stars within the core see a fluctuation of mass within the core radius. The perturbation Hamiltonian is:
\begin{equation}
\Delta H = \mathcal{C} \frac{\Omega_a}{\Omega_d} V_0 \frac{M_{\rm DM}}{M_\star} \cos \omega_{\rm osc} t \, , 
\end{equation}
where $\mathcal{C}\approx 0.3$~\cite{Veltmaat2018}. The time evolution of the semi-major axis is given by
\begin{equation}
\dot{a} = 2\mathcal{C} \frac{\Omega_a}{\Omega_d}\frac{M_{\rm DM}}{M_\star}a_0^2\omega_{\rm osc}\frac{\sin \omega_{\rm osc} t}{r} \, ,
\label{eqn:semi-major} 
\end{equation}
where $r(t)$ is the unperturbed orbit. We fix eccentricity $e=0.5$ and have verified that the results are not strongly dependent on this choice. 
%The time evolution of any orbital parameter $c_i(w,J)$ is given by
%\begin{equation}
%\dot{c}_i = \sum_j[c_i,c_j]\frac{\partial \Delta H}{\partial c_j} \, , 
%\end{equation}
%where $[\cdot,\cdot]$ is the Poisson bracket, and the sum runs over the set of six generalized phase space co-ordinates. Taking the $c_i$ co-ordinates to be given by the action-angle variables, except replacing $J_3$ with $a$ gives:

The solution for $a(t)$ is found by integrating Eq.~\eqref{eqn:semi-major} with initial condition $a(0)=a_0$. After one long cycle (the longer of $\tau_{\rm orb}$ and $\tau_{\rm osc}$) the evolution settles down into a new periodic state around a different value of $a$. We take the final value of $a$ to be the average over the period $\tau_{\rm av}=10\times {\rm Max}(\tau_{\rm orb},\tau_{\rm osc})$. Constraints are imposed by demanding the average orbit size does not double, $a_{\rm final}/a_0<2$. The perturbation analysis excludes four orbital resonances in the range $10\lesssim m_{22}\lesssim 50$, as shown in Fig.~\ref{fig:plot1}. Higher masses lead to small expansion of the star cluster, but with $M_{\rm DM}=330 M_\odot$ they are not significant. When $\tau_{\rm osc}<\tau_{\rm orb}$ (see Eq.~\ref{eqn:periods}) the soliton oscillations are adiabatic and do not affect the orbital parameters. Thus, FDM masses below the first resonance band, $m_a\lesssim 10^{-21}\text{ eV}$, if consistent with the formation of Eri-II, would also be consistent with the size and age of the star cluster.

\emph{Discussion:} Our results have placed strong constraints on FDM as a large fraction of the DM in an as-yet-unprobed high mass region. The formation of Eri II as a subhalo demands $m_a\gtrsim 0.8\times 10^{-21}\text{ eV}$ if FDM is all of the DM. The range $0.8\times 10^{-21}\text{ eV}\lesssim m_a\lesssim 10^{-19}\text{ eV}$ is further disfavoured by the observed size and age of the Eri II star cluster. In the small window $m_a=\mathcal{O}({\rm few})\times 10^{-21}\text{ eV}$ the diffusion approximation we have used is only partially applicable, however a series of resonances in this window further affect star cluster stability. 

The vanilla FDM model with no self-interactions is excluded by black hole superradiance for masses in the range $10^{-19}\text{ eV}\lesssim m_a\lesssim 10^{-16}\text{ eV}$~\cite{2011PhRvD..83d4026A,Stott:2018opm}. In addition to Eri II, Ref.~\cite{Brandt2016} considered the survival of ultra compact dwarfs with PBH DM. The bounds are equivalent to those from Eri II under observational and modelling uncertainties.  Ref.~\cite{Koushiappas:2017chw} found tighter constraints on light PBHs from mass segregation in Segue I. If our analysis were applied to this galaxy, the bound on FDM mass would be increased higher than $10^{-19}\text{ eV}$, deeper into the region disfavoured by superradiance. A complete study of dynamical constraints on FDM is desirable, but is unlikely to change the main conclusions of the present work.
%, and masses larger than $10^{-16}\text{ eV}$ are unlikely to have any observational signatures distinct from CDM on astrophysical scales. %The upper edge of the limit from Eri II meets the superradiance bound potentially excluding FDM entirely.

An FDM mass $m_a\lesssim 10^{-22}\text{ eV}$ is required for FDM to provide a resolution of the cusp-core problem in dSphs~\cite{2014NatPh..10..496S,2015MNRAS.451.2479M,Gonzales-Morales:2016mkl,2017MNRAS.468.1338C} (baryonic processes e.g. feedback are expected to play the dominant role~\cite{1996MNRAS.283L..72N,2012MNRAS.422.1231G,2014MNRAS.437..415D}). A dominant FDM component in the range $10^{-22}\text{ eV}$ to $10^{-21}\text{ eV}$ is in conflict with the Lyman alpha forest (though see Ref.~\cite{Hui2017}), and with the formation of low mass satellite galaxies like Eri II. 
%Constraints on this range could be weakened within modelling uncertainties (e.g. on stripping, and the temperature of the intergalactic medium), and 
In this range there are no additional constraints from star cluster heating in Eri II. Some authors have argued that $m_a\sim \mathcal{O}(\text{few})\times 10^{-22}\text{ eV}$ is favoured by density profiles of local dSphs~\cite{Broadhurst:2019fsl} and the Milky Way core~\cite{DeMartino:2018zkx}, and may even be favoured by an apparent turn over in the Hubble Frontiers Fields luminosity function~\cite{2018ApJ...862..156L}.

Thus there is a small window remaining for FDM with mass $\sim 10^{-21}\text{ eV}$. Such FDM will lie just below or in between the resonance bands of Eri II, suggesting that star cluster heating and resonances in other systems could be a new tool to search for FDM. %The survival of FDM in this window must be determined by dedicated simulations of star clusters inside oscillating FDM halos, and motivates a more detailed study of the stellar orbits in Eri II and other UFDs.

%Any FDM masses of astrophysical relevance outside of our small window require going beyond the non-interacting model to be consistent with observations. Assuming a UV completion of FDM as an ultralight axion with a cosine instanton potential, the strength of interactions is set by the decay constant, $f_a$. Lower values of $f_a$ increase the strength of axion self-interactions and relieve the constraints from black hole superradiance due to the Bosenova effect~\cite{2015PhRvD..91h4011A}. Lower values of $f_a$ also lead to tachyonic instabilities during structure formation~\cite{Cedeno:2017sou,Zhang:2017dpp} and relieve the Lyman-alpha forest bounds~\cite{Schive:2017biq,Leong:2018opi}. 

%The addition of significant self-interactions makes the phenomenology of FDM much more interesting: non-linear structure evolves differently~\cite{Desjacques:2017fmf}, and axion stars are more unstable to collapse and nova~\cite{Levkov:2016rkk,Helfer:2016ljl}. These phenomena could lead to as yet unexplored signatures of ultralight axions, but require simulations beyond those of vanilla FDM. Furthermore, at higher axion masses and lower decay constants the prospects for direct detection are greatly enhanced~\cite{Kim:2015yna,2018QS&T....3a4008G}.

\emph{Note added:} While this work was in preparation, Ref.~\cite{2018arXiv180907673B} appeared, which also derives relaxation effects produced by FDM halo fluctuations, applied to the cases of dynamical friction of very massive objects (satellites, supermassive black holes), and to the heating of early type galaxies.

\acknowledgements{This work made use of the open source packages \textsc{AstroPy}~\cite{astropy:2018} and \textsc{PyAstronomy}~\footnote{\url{https://www.hs.uni-hamburg.de/DE/Ins/Per/Czesla/PyA/PyA/index.html}}. DJEM is supported by the Alexander von Humboldt Foundation and the German Federal Ministry of Education and Research. }

\bibliographystyle{h-physrev3}

\bibliography{axion}

\begin{thebibliography}{10}

\bibitem{2005PhR...405..279B}
G.~{Bertone}, D.~{Hooper}, and J.~{Silk},
\newblock \physrep {\bf 405}, 279 (2005), hep-ph/0404175.

\bibitem{Aghanim:2018eyx}
Planck, N.~Aghanim {\em et~al.},
\newblock (2018), 1807.06209.
%%CITATION = ARXIV:1807.06209;%%

\bibitem{2015NatPh..11..245I}
F.~{Iocco}, M.~{Pato}, and G.~{Bertone},
\newblock Nature Physics {\bf 11}, 245 (2015), 1502.03821.

\bibitem{pdg}
Particle Data Group, K.~Olive {\em et~al.},
\newblock Chin. Phys. {\bf C38}, 090001 (2014).
%%CITATION = CHPHD,C38,090001;%%

\bibitem{hu2000}
W.~Hu, R.~Barkana, and A.~Gruzinov,
\newblock \prl {\bf 85}, 1158 (2000), astro-ph/0003365.
%%CITATION = ASTRO-PH/0003365;%%

\bibitem{2014MNRAS.437.2652M}
D.~J.~E. {Marsh} and J.~{Silk},
\newblock \mnras {\bf 437}, 2652 (2014), 1307.1705.

\bibitem{2016PhR...643....1M}
D.~J.~E. {Marsh},
\newblock \physrep {\bf 643}, 1 (2016), 1510.07633.

\bibitem{Hui2017}
L.~Hui, J.~P. Ostriker, S.~Tremaine, and E.~Witten,
\newblock \prd {\bf 95}, 043541 (2017), 1610.08297.

\bibitem{Brandt2016}
T.~D. Brandt,
\newblock The Astrophysical Journal {\bf 824}, L31 (2016), 1605.03665.

\bibitem{Li2016}
T.~S. Li {\em et~al.},
\newblock (2016), 1611.05052.

\bibitem{2014JCAP...02..019K}
A.~{Khmelnitsky} and V.~{Rubakov},
\newblock \jcap {\bf 2}, 019 (2014), 1309.5888.

\bibitem{Blas:2016ddr}
D.~Blas, D.~L. Nacir, and S.~Sibiryakov,
\newblock Phys. Rev. Lett. {\bf 118}, 261102 (2017), 1612.06789.
%%CITATION = ARXIV:1612.06789;%%

\bibitem{Aoki:2016kwl}
A.~Aoki and J.~Soda,
\newblock Int. J. Mod. Phys. {\bf D26}, 1750063 (2016), 1608.05933.
%%CITATION = ARXIV:1608.05933;%%

\bibitem{DeMartino:2017qsa}
I.~De~Martino {\em et~al.},
\newblock Phys. Rev. Lett. {\bf 119}, 221103 (2017), 1705.04367.
%%CITATION = ARXIV:1705.04367;%%

\bibitem{Boskovic:2018rub}
M.~Boskovic, F.~Duque, M.~C. Ferreira, F.~S. Miguel, and V.~Cardoso,
\newblock Phys. Rev. {\bf D98}, 024037 (2018), 1806.07331.
%%CITATION = ARXIV:1806.07331;%%

\bibitem{Abel:2017rtm}
C.~Abel {\em et~al.},
\newblock Phys. Rev. {\bf X7}, 041034 (2017), 1708.06367.
%%CITATION = ARXIV:1708.06367;%%

\bibitem{khlopov_scalar}
M.~Khlopov, B.~Malomed, and I.~Zeldovich,
\newblock \mnras {\bf 215}, 575 (1985).
%%CITATION = MNRAA,215,575;%%

\bibitem{Hlozek:2017zzf}
R.~Hlozek, D.~J.~E. Marsh, and D.~Grin,
\newblock Mon. Not. Roy. Astron. Soc. {\bf 476}, 3063 (2018), 1708.05681.
%%CITATION = ARXIV:1708.05681;%%

\bibitem{2015MNRAS.450..209B}
B.~{Bozek}, D.~J.~E. {Marsh}, J.~{Silk}, and R.~F.~G. {Wyse},
\newblock \mnras {\bf 450}, 209 (2015), 1409.3544.

\bibitem{2016ApJ...818...89S}
H.-Y. {Schive}, T.~{Chiueh}, T.~{Broadhurst}, and K.-W. {Huang},
\newblock \apj {\bf 818}, 89 (2016), 1508.04621.

\bibitem{2016JCAP...04..012S}
A.~{Sarkar} {\em et~al.},
\newblock \jcap {\bf 4}, 012 (2016), 1512.03325.

\bibitem{2017PhRvD..95h3512C}
P.~S. {Corasaniti}, S.~{Agarwal}, D.~J.~E. {Marsh}, and S.~{Das},
\newblock \prd {\bf 95}, 083512 (2017), 1611.05892.

\bibitem{Irsic:2017yje}
V.~Irsic, M.~Viel, M.~G. Haehnelt, J.~S. Bolton, and G.~D. Becker,
\newblock Phys. Rev. Lett. {\bf 119}, 031302 (2017), 1703.04683.
%%CITATION = ARXIV:1703.04683;%%

\bibitem{Armengaud:2017nkf}
E.~Armengaud, N.~Palanque-Delabrouille, C.~Y{\`{e}}che, D.~J.~E. Marsh, and
  J.~Baur,
\newblock Mon. Not. Roy. Astron. Soc. {\bf 471}, 4606 (2017), 1703.09126.
%%CITATION = ARXIV:1703.09126;%%

\bibitem{Nori:2018pka}
M.~Nori, R.~Murgia, V.~Ir?i?, M.~Baldi, and M.~Viel,
\newblock (2018), 1809.09619.
%%CITATION = ARXIV:1809.09619;%%

\bibitem{2017MNRAS.465..941D}
X.~{Du}, C.~{Behrens}, and J.~C. {Niemeyer},
\newblock \mnras {\bf 465}, 941 (2017), 1608.02575.

\bibitem{2014NatPh..10..496S}
H.-Y. {Schive}, T.~{Chiueh}, and T.~{Broadhurst},
\newblock Nature Physics {\bf 10}, 496 (2014), 1406.6586.

\bibitem{Veltmaat2018}
J.~Veltmaat, J.~C. Niemeyer, and B.~Schwabe,
\newblock Physical Review D {\bf 98}, 043509 (2018), 1804.09647.

\bibitem{2018arXiv180904744C}
B.~V. {Church}, J.~P. {Ostriker}, and P.~{Mocz},
\newblock ArXiv e-prints  (2018), 1809.04744.

\bibitem{Amorisco:2018dcn}
N.~C. Amorisco and A.~Loeb,
\newblock (2018), 1808.00464.
%%CITATION = ARXIV:1808.00464;%%

\bibitem{Levkov:2018kau}
D.~G. Levkov, A.~G. Panin, and I.~I. Tkachev,
\newblock (2018), 1804.05857.
%%CITATION = ARXIV:1804.05857;%%

\bibitem{2014PhRvL.113z1302S}
H.-Y. {Schive} {\em et~al.},
\newblock Physical Review Letters {\bf 113}, 261302 (2014), 1407.7762.

\bibitem{Du:2016aik}
X.~Du, C.~Behrens, J.~C. Niemeyer, and B.~Schwabe,
\newblock \prd {\bf 95}, 043519 (2017), 1609.09414.
%%CITATION = ARXIV:1609.09414;%%

\bibitem{Bar:2018acw}
N.~Bar, D.~Blas, K.~Blum, and S.~Sibiryakov,
\newblock \prd {\bf 98}, 083027 (2018), 1805.00122.
%%CITATION = ARXIV:1805.00122;%%

\bibitem{2012JCAP...02..011L}
V.~{Lora}, J.~{Maga{\~n}a}, A.~{Bernal}, F.~J. {S{\'a}nchez-Salcedo}, and E.~K.
  {Grebel},
\newblock \jcap {\bf 2}, 11 (2012), 1110.2684.

\bibitem{2015MNRAS.451.2479M}
D.~J.~E. {Marsh} and A.-R. {Pop},
\newblock \mnras {\bf 451}, 2479 (2015), 1502.03456.

\bibitem{2017MNRAS.468.1338C}
S.-R. {Chen}, H.-Y. {Schive}, and T.~{Chiueh},
\newblock \mnras {\bf 468}, 1338 (2017), 1606.09030.

\bibitem{Gonzales-Morales:2016mkl}
A.~X. Gonz\'{a}lez-Morales, D.~J.~E. Marsh, J.~Pe\~{n}arrubia, and L.~A.
  Ure\~{n}a L\'{o}pez,
\newblock \mnras {\bf 472}, 1346 (2017), 1609.05856.
%%CITATION = ARXIV:1609.05856;%%

\bibitem{2016MNRAS.460.4397C}
E.~{Calabrese} and D.~N. {Spergel},
\newblock \mnras {\bf 460}, 4397 (2016), 1603.07321.

\bibitem{Robles:2018fur}
V.~H. Robles, J.~S. Bullock, and M.~Boylan-Kolchin,
\newblock (2018), 1807.06018.
%%CITATION = ARXIV:1807.06018;%%

\bibitem{2018arXiv180708153D}
I.~{De Martino}, T.~{Broadhurst}, S.-H.~H. {Tye}, T.~{Chiueh}, and H.-Y.
  {Schive},
\newblock ArXiv e-prints  (2018), 1807.08153.

\bibitem{2018ApJ...862..156L}
E.~{Leung} {\em et~al.},
\newblock \apj {\bf 862}, 156 (2018), 1806.07905.

\bibitem{2016ApJ...824L..14C}
D.~{Crnojevi{\'c}} {\em et~al.},
\newblock \apjl {\bf 824}, L14 (2016), 1604.08590.

\bibitem{Contenta2018}
F.~Contenta {\em et~al.},
\newblock Monthly Notices of the Royal Astronomical Society {\bf 476}, 3124
  (2018), 1705.01820.

\bibitem{XiaolongThesis}
X.~{Du},
\newblock {\em {PhD Thesis}} ({Georg-August Universit\"{a}t, G\"{o}ttingen},
  2018).

\bibitem{Du:2016zcv}
X.~Du, C.~Behrens, and J.~C. Niemeyer,
\newblock Mon. Not. Roy. Astron. Soc. {\bf 465}, 941 (2017), 1608.02575.
%%CITATION = ARXIV:1608.02575;%%

\bibitem{Du:2018zrg}
X.~Du, B.~Schwabe, J.~C. Niemeyer, and D.~Burger,
\newblock Phys. Rev. {\bf D97}, 063507 (2018), 1801.04864.
%%CITATION = ARXIV:1801.04864;%%

\bibitem{2012NewA...17..175B}
A.~J. {Benson},
\newblock New Astronomy {\bf 17}, 175 (2012), 1008.1786.

\bibitem{Schneider:2014rda}
A.~Schneider,
\newblock Mon. Not. Roy. Astron. Soc. {\bf 451}, 3117 (2015), 1412.2133.
%%CITATION = ARXIV:1412.2133;%%

\bibitem{2016JCAP...04..059S}
A.~{Schneider},
\newblock \jcap {\bf 4}, 059 (2016), 1601.07553.

\bibitem{2017JCAP...11..046M}
R.~{Murgia}, A.~{Merle}, M.~{Viel}, M.~{Totzauer}, and A.~{Schneider},
\newblock \jcap {\bf 11}, 046 (2017), 1704.07838.

\bibitem{Eby:2018zlv}
J.~Eby, M.~Leembruggen, P.~Suranyi, and L.~C.~R. Wijewardhana,
\newblock (2018), 1805.12147.
%%CITATION = ARXIV:1805.12147;%%

\bibitem{BinneyTremaine2008}
J.~{Binney} and S.~{Tremaine},
\newblock {\em {Galactic Dynamics: Second Edition}} ({Princeton University
  Press}, 2008).

\bibitem{El-Zant2016}
A.~A. El-Zant, J.~Freundlich, and F.~Combes,
\newblock Monthly Notices of the Royal Astronomical Society {\bf 461}, 1745
  (2016), 1603.00526.

\bibitem{Osterbrock1952}
D.~E. {Osterbrock},
\newblock \apj {\bf 116}, 164 (1952).

\bibitem{2018arXiv180907673B}
B.~{Bar-Or}, J.-B. {Fouvry}, and S.~{Tremaine},
\newblock ArXiv e-prints  (2018), 1809.07673.

\bibitem{Veltmaat2019}
J.~Veltmaat, J.~C. Niemeyer, and B.~Schwabe,
\newblock in preparation  (2019).

\bibitem{goldstein1980classical}
H.~Goldstein,
\newblock {\em Classical Mechanics} (Addison-Wesley, 1980).

\bibitem{2011PhRvD..83d4026A}
A.~{Arvanitaki} and S.~{Dubovsky},
\newblock \prd {\bf 83}, 044026 (2011), 1004.3558.

\bibitem{Stott:2018opm}
M.~J. Stott and D.~J.~E. Marsh,
\newblock Phys. Rev. {\bf D98}, 083006 (2018), 1805.02016.
%%CITATION = ARXIV:1805.02016;%%

\bibitem{Koushiappas:2017chw}
S.~M. Koushiappas and A.~Loeb,
\newblock Phys. Rev. Lett. {\bf 119}, 041102 (2017), 1704.01668.
%%CITATION = ARXIV:1704.01668;%%

\bibitem{1996MNRAS.283L..72N}
J.~F. {Navarro}, V.~R. {Eke}, and C.~S. {Frenk},
\newblock \mnras {\bf 283}, L72 (1996), astro-ph/9610187.

\bibitem{2012MNRAS.422.1231G}
F.~{Governato} {\em et~al.},
\newblock \mnras {\bf 422}, 1231 (2012), 1202.0554.

\bibitem{2014MNRAS.437..415D}
A.~{Di Cintio} {\em et~al.},
\newblock \mnras {\bf 437}, 415 (2014), 1306.0898.

\bibitem{Broadhurst:2019fsl}
T.~Broadhurst, I.~de~Martino, H.~N. Luu, G.~F. Smoot, and S.~H.~H. Tye,
\newblock (2019), 1902.10488.
%%CITATION = ARXIV:1902.10488;%%

\bibitem{DeMartino:2018zkx}
I.~De~Martino, T.~Broadhurst, S.~H.~H. Tye, T.~Chiueh, and H.-Y. Schive,
\newblock (2018), 1807.08153.
%%CITATION = ARXIV:1807.08153;%%

\bibitem{astropy:2018}
A.~M. {Price-Whelan} {\em et~al.},
\newblock aj {\bf 156}, 123 (2018).

\end{thebibliography}

\end{document}